
\headline={\ifnum\pageno=1\firstheadline\else
\ifodd\pageno\rightheadline \else\leftheadline\fi\fi}
\def\firstheadline{\hfil}
\def\rightheadline{\hfil}
\def\leftheadline{\hfil}
	\footline={\ifnum\pageno=1\firstfootline\else\otherfootline\fi}
\def\firstfootline{\rm\hss\folio\hss}
\def\otherfootline{\hfil}

\font\twelvebf=cmbx10 scaled\magstep 1
\font\twelverm=cmr10 scaled\magstep 1
\font\twelveit=cmti10 scaled\magstep 1

\font\tenrm=cmr10
\font\tenit=cmti10

\font\ninerm=cmr9

\parindent=1.5pc
\hsize=6.0truein
\vsize=8.5truein
\nopagenumbers
\def\l{\lambda}\def\s{\scriptstyle}\def\ss{\scriptscriptstyle}
\def\f{\phi}\def\ff{\f^4}\def\k{\kappa}\def\p{\partial}\def\o{\over}
\def\e{\varepsilon}\def\gf{\gamma_{\f}}
\def\gft{\gamma_{\f^2}} \def\ra{\rightarrow}\def\disp{\displaystyle}
\def\i{\infty} \def\frac#1#2{{{#1}\over{#2}}}
 \def\de{d_{\ss eff}}\def\fb{\bar\f}

\hfill{DIAS-STP-93-25; Oct.'93}
\vskip 0.1truein
\centerline{\twelvebf  HEATING FIELD THEORY THE
\footnote{\dag}
{\ninerm To be published in ``Proceedings of the 3rd Workshop on Thermal
Field Theories and their Applications'', Banff, Canada 1993.}
}
\baselineskip=22pt
\centerline{\twelvebf  ``ENVIRONMENTALLY FRIENDLY''  WAY!}
\vglue 0.7cm
\centerline{\tenrm M.A. VAN EIJCK}
\baselineskip=13pt
\centerline{\tenit Inst. for Theor. Physics, University of Amsterdam,
Valckenierstraat 65}
\baselineskip=12pt
\centerline{\tenit NL-1018 XE Amsterdam, Netherlands}
\vglue 0.3cm
\centerline{\tenrm DENJOE O'CONNOR}
\centerline{\tenrm and}
\centerline{\tenrm C.R. STEPHENS}
\baselineskip=13pt
\centerline{\tenit  D.I.A.S., 10 Burlington Road, Dublin 4, Ireland}
\vglue 0.8cm
\centerline{\tenrm ABSTRACT}
\vglue 0.3cm
{\rightskip=3pc
 \leftskip=3pc
 \tenrm\baselineskip=12pt\noindent
We discuss how to implement an ``environmentally friendly''
renormalization in the context of finite temperature field theory.
Environmentally friendly renormalization provides a method for
interpolating between the different effective field theories which
characterize different asymptotic regimes.  We give explicit two loop
Pad\'e resummed results for $\l\ff$ theory for $T>T_c$. We examine the
implications for non-Abelian gauge theories.
\vglue 0.6cm}

\vfil
\twelverm
\baselineskip=14pt
\leftline{\twelvebf 1. Introduction}
\vglue 0.4cm
Many of you will no doubt, at some time or other, have tried to
understand the infrared structure of finite temperature field theory.
There are many techniques available: ``daisy'' resummations, two
particle irreducible effective action, $1/N$ expansions, $\varepsilon$
expansions etc.. We suspect however that you haven't been completely
satisfied with any of them. In this paper we will discuss a methodology
which offers significant advantages over other techniques ---
``environmentally friendly'' renormalization.  The approach was
initiated by O'Connor and Stephens$^{\s 1}$, and further developed in
subsequent papers$^{\s 2}$. A full and recent account$^{\s 3}$ of
``environmentally friendly'' methods is available, here, because of
space constraints, we can only offer a brief summary Let us first
remind ourselves about some of the main problems of finite temperature
field theory. When one heats a theory up one is basically using the
temperature parameter as a means of changing ``scale''. Physical
systems generically exhibit effective degrees of freedom (EDOF) which
are radically different at different scales.  The keyboard at which one
of us is writing this paper for instance has 86 keys which can be
depressed or undepressed giving rise to 86 EDOF.  Obviously at scales
$\sim 10^{-6}{\rm cm}$ the EDOF are quite different. It would be quite
inappropriate to describe the typing process in terms of the atoms
which make up the keys. Equally, a description of microscopic physics
in terms of typewriter keys is inappropriate. Additionally if I raised
the temperature of the typewriter eventually the natural degrees of
freedom would be gas molecules. The example might seems somewhat
facetious but it has a serious point. In QCD or the electroweak theory
the EDOF at temperatures $\sim 1{\rm Mev}$ are very different to those
at temperatures $\sim 300{\rm Mev}$.  More prosaically $\l\ff$ theory,
a paradigm for the Higgs sector of the standard model, exhibits very
different degrees of freedom at very high and very low temperatures.

The basic problem of finite temperature field theory is to
quantitatively describe systems which exhibit very different degrees of
freedom at different ``scales'' --- temperatures. In the next section
we will outline the basis of environmentally friendly renormalization
wherein such systems can be examined. In section 3 we will discuss
exactly what one means by a high temperature system. In section 4 we
will present results for $\l\ff$ theory and in section 5 discuss gauge
theories. Finally in section 6 we'll draw some conclusions.

\vglue 0.6cm
\leftline{\twelvebf 2. ``Environmentally Friendly'' Renormalization}
\vglue 0.4cm
Field theories describe systems that have many degrees of freedom. In
such systems fluctuations inevitably play a very important role.
Renormalization judiciously applied provides a method of capturing the
effects of fluctuations in the parameters of the theory. The bare
parameters with which one begins a field theory calculation are
suitable for describing systems in the absence of fluctuations and
describe in a meaningful way the EDOF of the system under such
circumstances. Fluctuations change the nature of the EDOF to such an
extent that the bare parameters become totally unsuited to describing
them. One gets round this problem by using the freedom to reparametrize
the system, transforming to a set of parameters, the renormalized ones,
that offer a more faithful representation of the EDOF.

Particle physicists have traditionally seen the UV divergences
seemingly inherent in a continuum description as the reason for
renormalization.  There is nothing inherent in the theory of
renormalization itself to warrant this. Large fluctuation effects can
in priciple originate anywhere in the spectrum. The canonical words:
``divergences'', ``strong-coupling'' and ``non-perturbative'' are
almost the inevitable consequence of trying to describe a system with
strong fluctuations in terms of an inappropriate set of parameters.
What one requires is a renormalization/reparametrization to a set of
parameters suitable for describing the system at the scales of
interest.

In the case of finite temperature field theory the EDOF are temperature
dependent. It therefore seems sensible to implement a renormalization
which is temperature dependent in order that one might track the
evolving nature of the EDOF as a function of temperature. The
imposition of a temperature independent renormalization yields a very
badly behaved perturbative series. In the high temperature limit
perturbation theory breaks down and (IR) divergences appear. This
breakdown is telling us that the renormalized parameters being used are
totally inadequate for describing the high temperature limit, being
those appropriate for $T=0$, e.g. $T=0$ masses and coupling constants.
If one thinks of renormalization as a coarse graining procedure, then
when one changes renormalization scale, i.e. implements an RG
transformation, using $T=0$ renormalized parameters, one is effectively
coarse graining $T=0$ EDOF.  As one heats the theory up the disparity
between the ``true'' EDOF of the system and those effectively being
coarse grained becomes greater, hence the breakdown in perturbation
theory.

All the above undesireable effects can be avoided by implementing a
more suitable RG. We will show this more explicitly in the next
section. Besides temperature there are many other parameters that could
induce  changes in the EDOF: e.g. electric/magnetic fields, spacetime
geometry, anisotropic interactions etc. We dub such parameters
``environmental'' as we are generically trying to describe fluctuations
in a field theory in an ``environment'' described by these parameters.
Thus in QED changing a background magnetic field changes the
environment in which the electrons and photons see themselves. As the
EDOF are sensitive to the environment, we will call an RG which tracks
the changing nature of them as the environment changes
``environmentally friendly''

\vglue 0.6cm
\leftline{\twelvebf 3. The Environmentally Friendly Way}
\vglue 0.4cm
In this section we will consider finite temperature $\l\ff$ theory. We
will be sketchy leaving the reader to get more details from the
papers$^{\s 1,2,3}$. We determine our renormalized parameters via the
normalization conditions
$$\Gamma^{(2)}(k=0,m^2=\k^2,\l,T,\k)=\k^2\eqno(1)$$
$${\p\Gamma^{(2)}\o\p k^2}(k,m^2=\k^2,\l,T,\k)|_{\ss k=0}=1\eqno(2)$$
$$\Gamma^{(4)}(k=0,m^2=\k^2,\l,T,\k)=\l\eqno(3)$$
$$\Gamma^{(2,1)}(k=0,m^2=\k^2,\l,T,\k)=1\eqno(4)$$
$\k$ being an arbitrary renormalization scale. Using these conditions,
and implementing a [2,1] Pad\'e resummation of the two loop Wilson functions
we obtain
$$\beta(h,\tau)=-\e(\tau) h+{h^2\o 1+
4\left({(5N+22)\o {(N+8)}^{2}}f_{1}(\tau)
-{(N+2)\o{(N+8)}^{2}}f_{2}(\tau)\right) h}\eqno(5)$$
$$\gft(h,\tau)={(N+2)\o (N+8)}{h\o1+6{1\o{(N+8)}}\bigl(f_{1}(\tau)
-{1\o 3}f_{2}(\tau)\bigr) h}\eqno(6)$$ and
$$\gf(h,\tau)=2{(N+2)\o{(N+8)^2}}f_2(\tau)h^2\eqno(7)$$
where $\varepsilon(\tau)=1+\tau{d\over d\tau}{\rm ln}
({\disp\sum_n}m^{-3})$, $\tau={T\over\kappa}$
$$f_1(\tau) = 2{{\disp\sum_{n_1,n_2}}({1\over m_1^3}({1\over M}-{1\over2m_2})
+{1\over m_1M^2}({1\over m_1}+{2\over m_2}))
\over({\disp\sum_{n}}{1\over m^{3}})^{2}}\qquad \hbox{ and }\quad
f_2(\tau)=4{{\disp\sum_{n_1,n_2}}{1\over M^3m_1}
\over({\disp\sum_{n}}{1\over m^3})^{2}}$$
with
$m_i=(1+{4\pi^2n^2_{i}\over \tau^2})^{\s\frac12}$, $m_{12}=(1+{4\pi^2\over
\tau^2}(n_1+n_2)^2)^{\s\frac12}$, $M=m_1+m_2+m_{12}$. In Eq.'s~(5-7) the
coupling $h$, or floating coupling$^{\s 1}$, is defined via the relation
$h=a_2({T\o\k})\l(\k)$, where $a_2$ is the coefficient of $\l^2$ in
$\beta(\l)$. Eq.'s~(1) and (2) imply that $\k$ is the inverse finite
temperature screening length $m(T)$, thus the Wilson functions depend on
$T\o m(T)$ and $h$.

Eq.~(5) exhibits more than one fixed point. As ${T\o m(T)}\ra0$,
one obtains the
Gaussian fixed point as expected in four dimensions. As $T\ra T_c$,
i.e. $m(T)\ra0$ one finds a non-trivial fixed point, $h=1.732$ for
$N=1$, for instance. The value of the fixed point and the corresponding
critical exponents are in exact agreement with corresponding two loop
Pad\'e resummed results$^{\s 4}$ in three dimensional critical
phenomena. Of course we could have assumed that our theory was three
dimensional near the phase transition but why do that when we can
derive it. We can see the complete crossover between four and three
dimensional behaviour as $T\o m(T)$ varies between $0$ and $\i$ in a
perturbatively controllable fashion.  One can think of
$\de=4-\varepsilon({T\o m(T)})$,
which interpolates between 4 and 3 when $T\o m(T)$ varies between $0$ and
$\i$, as a measure of the effective dimension of the system.  Near
$T=0$ the EDOF of the problem are 4 dimensional and near $T=T_c$, 3
dimensional. Between these two extremes they are neither 4 nor 3
dimensional.  The power of our approach is that we have implemented a
reparametrization which is temperature dependent in such a fashion that
it tracks the evolving nature of the EDOF between the 4 and 3
dimensional limits. In the figure we present a plot of $\gf({T\o k})$
versus $\ln{T\o k}$, where the solution of $\k{d\l\o d\k}=\beta$
has been used. Its
physical significance derives from the fact that at $T=T_c$,
$$G^{(2)}(k,T)\sim {e^{\int^k \gf}\o k^2}\eqno(8)$$ where $k$ is the
spatial momentum.

The parameter with which we are investigating the crossover here is the
finite temperature screening length. Though it is a very natural parameter
in the real world often we only have access to the zero temperature
parameters. One would therefore like to know how to describe the
crossover in terms of them. Here we performed a renormalization of the
system at the physical temperature $T$ and used the screening length as
a running scale. We could have renormalized at a completely arbitrary
value of the temperature instead$^{\s 5}$. In this
case we would be running the environment itself. By so doing however
one can relate finite temperature quantities to zero temperature
quantities quite easily. More will be said about such schemes in
another article in these Proceedings.

\vglue 0.6cm
\leftline{\twelvebf 4. ``Watching the daisies grow''}
\vglue 0.4cm
One of the most oft used methods of treating IR problems in finite
temperature field theory, and certainly in $\l\ff$ theory, is to resum
the daisy diagrams on the basis that they are the most IR divergent as
temperature increases.  It is important to have an intuitive
understanding of the approximation procedure one is implementing. We
have motivated our methods in terms of a reparametrization which tracks
the EDOF of the system as the environment changes. If perturbation
theory works you can be fairly sure you are tracking the EDOF well. If
it breaks down the opposite is true.

When one starts with $\l\ff$ theory at $T=0$, with a zero temperature
mass $m$, one finds that when $T\gg m\ (T>T_c)$ perturbation theory
breaks down due to the presence of a large thermally induced mass
$m(T)=\l^{\ss {1\o2}}T$.  Perturbation theory is breaking down because
one is trying to use small mass EDOF to describe a system where the
true EDOF are very massive relative to the scale $m$. By resumming the
daisies one is effectively expanding around a theory of mass  $m(T)$ as
opposed to a mass $m$. Once this is done one finds that perturbation
theory works. So, obviously something was done correctly. But what is
being described? Certainly not the vicinity of a second order or weakly
first order phase transition where $m(T)\ll T$. Daisy resummation
methods alone cannot be used to describe such a regime because
the physical characteristic of the regime, $m(T)\ll T$, is totally
different to that of the  daisy resummation, $m\ll T$. When one starts
at $T=0$ with $\fb=0$ and heats the theory up, then one drives the
theory away from a phase transition not towards it. What regime is far
away from the critical regime in critical phenomena? The mean field
regime, and it is precisely this regime which the daisy resummation is
suitable for. Once the large mass shift has been accounted for, one is
in a regime where IR fluctuations are strongly suppressed by the
effects of the large mass --- hence loop corrections become unimportant.
This is the one regime where it is not necessary to use RG methods
(though certainly RG methods can also describe this regime) because of
the fact that fluctuations are unimportant.

\vglue 0.6cm
\leftline{\twelvebf 4. Gauge Theories}
\vglue 0.4cm
We have tried to point out so far the intuitive basis of our
methodology and used $\l\ff$ theory as an interesting test ground for
it (besides the Higgs sector it can also successfully describe in one
guise or another a myriad of physical systems in statistical
physics$^{\s 3}$). We are interested in investigating how the EDOF of a
physical system are sensitive to the environment they ``feel''
themselves to be in. If we consider non-trivial background fields then
they too form part of the environment in which the fluctuations of the
system exist. These background fields need not be homogeneous but could
represent such interesting objects as instantons, solitons, monopoles,
vortices  etc.  As is well known such non-trivial solutions of the
classical field equations play a very important role in gauge theories.
Hence we would expect the environment in theories with gauge fields to
be very rich and complex. Gauge theories are therefore archetypal
crossover problems --- in QCD the low energy EDOF are entirely
different from the high energy ones. The growth of the QCD coupling in
the IR is symptomatic of the fact that the EDOF are no longer quarks
and gluons but hadrons and mesons. In principle this is no different
than the breakdown of perturbation theory in the IR limit of finite
temperature $\l\ff$ theory. The crucial difference is that in the
latter we know exactly how to describe the effects of temperature as an
environmental parameter.  We know much less about how to describe the
environmental effects of the QCD vacuum.

So what can we say about finite temperature gauge fields? The first
important fact is that there is a strong anisotropy in gauge theories
as to the effects of temperature. The electric and magnetic sectors
react in very different ways. In abelian gauge theories the electric
sector acquires a thermal mass $\sim eT$, while the magnetic sector
remains massless. The electric sector is then very akin to $\l\ff$
theory for $T\gg m$. Fluctuations acquire a large mass and essentially
become unimportant leading to mean field like behaviour. The magnetic
sector remains massless throughout and one might be tempted to think
that it is purely described by a three dimensional gauge field.
However, there is an interesting crossover between three and four
dimensional massless behaviour as a function of $T\o k$ where $k$ is
the typical momentum scale of the magnetic process under study. In all
these cases an environmentally friendly RG can proffer a reliable
description. That is not to say that there are no subtleties to beware
of. In scalar electrodynamics for instance vortices can play an
important role in lower dimensions, and one would consequently expect
to see a breakdown in renormalized perturbation theory if this fact is
ignored. The latter indicates that the original model for the
environment (i.e. ignoring vortices) was inadequate.

Non-Abelian gauge theories are even more complicated. As in the abelian
case the electric sector acquires a thermal mass $\sim gT$ and leads to
a mean field regime.  The magnetic sector however also picks up a
thermal mass, but only at the two loop level. If one works to one loop
in RG improved perturbation theory, where the only environmental
variable accounted for is temperature, one finds that due to the
absence of an IR cutoff one is driven into a strong coupling
regime$^{\s 6}$. Thus the simple prescription that works for $\l\ff$
theory is inadequate.  The moral is clear: thermally corrected quarks
and gluons are not the appropriate EDOF. The effect of the temperature
is to drive the theory into the confining regime and unfortunately the
model of the environment we are using is not sophisticated enough to
account properly for the effects of the QCD vacuum. One might argue
that including two loop effects in the thermal RG would alleviate this
problem.  This seems unlikely. Not only must a magnetic mass be
introduced but it must increase with temperature sufficiently fast so
as to act as an efficient IR cutoff at all temperatures.

\vglue 0.6cm
\leftline{\twelvebf Conclusions}
\vglue 0.4cm
In this short note we have described the essence of  environmentally
friendly renormalization which provides a method for capturing the
crossover from one effective field theory to another.  Effective field
theory focusses on a narrow band of energy scales within which the EDOF
don't qualitatively change. The power of our method is its ability to
capture the large qualitative changes in EDOF inherent in the crossover
between effective field theories.  The example we have concentrated on
here is finite temperature field theory, and in particular $\l\ff$
theory, where for ${T\o m(T)}\sim 0$ the effective field theory is four
dimensional, and for ${T\o m(T)}\gg1$ the effective field theory is
three dimensional.  We presented two loop Pad\'e resummed expressions
for the Wilson functions which capture fully this crossover as
exemplified in the figure. We also discussed briefly some of the
subtleties inherent in describing finite temperature gauge fields, a
subject we will return to in the future.

\vglue 0.6cm
\leftline{\twelvebf 6. References}
\vglue 0.4cm
\medskip
\itemitem{1.} Denjoe O'Connor and C.R. Stephens, {\twelveit Nucl. Phys.}
{\twelvebf B360} (1991) 297.
\itemitem{2.} Denjoe O'Connor, C.R. Stephens and F. Freire, {\twelveit Class.
Quan. Grav.} {\twelvebf 23} (1993) S243; Denjoe O'Connor, C.R. Stephens and F.
Freire,
{\twelveit Mod. Phys. Lett.} {\twelvebf A25} (1993) 1779; F. Freire and C.R.
Stephens,
{\twelveit Zeit. Phys.} {\twelvebf C} (to be published).
\itemitem{3.} Denjoe O'Connor and C.R. Stephens, ``Environmentally Friendly
Renormalization''; Preprint THU-93/14, DIAS-STP-93-19; (to be published in
{\twelveit
Int. Jou. Mod. Phys.} {\twelvebf A}.
\itemitem{4.} G.A. Baker, B.G. Nickel and D.I. Meiron, {\twelveit Phys. Rev.}
{\twelvebf B17} (1978) 1365.
\itemitem{5.} H. Matsumoto, Y. Nakano and H. Umezawa, {\twelveit Phys. Rev.}
{\twelvebf D29} (1984) 1116.
\itemitem{6.} M.A. van Eijck, C.R. Stephens and Ch.G. van Weert, ``Temperature
Dependence of the QCD Coupling'' ; Preprint THU-93/08, ITFA-93-11.

\bye